\documentclass[10pt]{iopart}
\usepackage{graphicx}
\usepackage{amssymb}
\usepackage{setstack}

\begin{document}
\title{Ferrimagnetism and disorder in epitaxial Mn$_{2-x}$Co$_x$VAl thin films} 

\author{Markus Meinert, Jan-Michael Schmalhorst, and G\"unter Reiss}
\address{Department of Physics, Bielefeld University,
33501 Bielefeld, Germany}
\ead{meinert@physik.uni-bielefeld.de}

\author{Elke Arenholz}
\address{Advanced Light Source, Lawrence Berkeley National Laboratory, CA 94720, USA}
\date{\today}

\begin{abstract}
The quaternary full Heusler compound Mn$_{2-x}$Co$_x$VAl with $x=1$ is predicted to be a half-metallic antiferromagnet. Thin films of the quaternary compounds with $x=0\dots2$ were prepared by DC and RF magnetron co-sputtering on heated MgO (001) substrates. The magnetic structure was examined by x-ray magnetic circular dichroism and the chemical disorder was characterized by x-ray diffraction. Ferrimagnetic coupling of V to Mn was observed for Mn$_2$VAl ($x = 0$). For $x = 0.5$, we also found ferrimagnetic order with V and Co antiparallel to Mn. The observed reduced magnetic moments are interpreted with the help of band structure calculations in the coherent potential approximation. Mn$_2$VAl is very sensitive to disorder involving Mn, because nearest-neighbor Mn atoms couple anti-ferromagnetically. Co$_2$VAl has B2 order and has reduced magnetization. In the cases with $ x\geq 0.9$ conventional ferromagnetism was observed, closely related to the atomic disorder in these compounds.
\end{abstract}

\submitto{\JPD}

\maketitle

\section{Introduction}

Half-metallic fully compensated ferrimagnets (HMFi), which are also known as half-metallic antiferromagnets, attracted large interest during the past years. A material with this property would exhibit full spin polarization at the Fermi level, but the magnetization would be effectively zero. It was first predicted for Mn and In doped FeVSb \cite{vanLeuken95}. Among others, La$_2$VMnO$_6$ and related double perovskites \cite{Pickett98}, and certain diluted magnetic semiconductors have been later predicted to be half-metallic antiferromagnets as well \cite{Akai06}. This is interesting for technological applications, e.g. for spin transfer torque switching, which depends on the magnetization and the spin polarization of the material to be switched.

Galanakis \textit{et al.} pointed out that it may be possible to synthesize a HMFi by substituting Co for Mn in the Heusler compound Mn$_2$VAl \cite{Galanakis07}. Mn$_2$VAl is a (potentially half-metallic) ferrimagnet with antiparallel coupling of Mn and V moments and a total moment of -2\,$\mu_B$ per formula unit. The high Curie temperature of 760\,K makes it interesting for practical applications. Numerous experimental \cite{Kawakami81,Yoshida81,Itoh83,Nakamichi83,Jiang01} and theoretical \cite{Ishida84,Weht99,Sasioglu05,Oezdogan06,Chioncel09} studies are found in the literature. Following the Slater-Pauling rule for Heusler compounds, $m = N_\mathrm{V} - 24$ \cite{Galanakis02}, the magnetic moment $m$ is to be taken as negative, because the number of valence electrons $N_\mathrm{V}$ is 22. Thus, by adding effectively two electrons per unit cell, the magnetization should vanish. This can be achieved by substituting one Mn with one Co atom, which has two additional electrons. \textit{Ab initio} simulations were carried out on this system in the L2$_1$ structure with Mn and Co randomly spread across the Wyckoff 8c sites and V and Al on the 4a and 4b sites. Indeed, a HMFi is found with magnetic moments of: -1.388 (Mn), 0.586 (Co), 0.782 (V), 0.019 (Al) \cite{Galanakis07}. It was shown by Luo \textit{et al.} that the site occupation preference in Mn$_2Y$Al depends on the number of valence electrons of $Y$: if it is lower than the one of Mn, $Y$ would preferentially occupy the 4a/b sites, but if it is higher, $Y$ would rather occupy the 8c sites together with Mn, changing the structure to the Hg$_2$CuTi type \cite{Luo08}. Accordingly, one can expect an occupation as proposed by Galanakis \textit{et al.} in Mn$_{2-x}$Co$_x$VAl (MCVA).

For many practical applications it is necessary to prepare thin films of the magnetic materials. Therefore one has to find suitable deposition techniques and optimize the parameters. The parent compounds Mn$_2$VAl and Co$_2$VAl \cite{Ziebeck74, Kanomata10} have been successfully synthesized in the bulk and epitaxial growth of Mn$_2$VAl films with L2$_1$ ordering on MgO (001) single crystals was also demonstrated \cite{Kubota09,Klaer10}. In this paper we present experimental results on the structural and magnetic properties of epitaxial Mn$_{2-x}$Co$_x$VAl thin films.

\section{Methods}
\subsection{Experimental details}

The samples were deposited using a UHV co-sputtering system equipped with five DC and two RF 3" magnetron sputtering sources, arranged in a confocal sputter-up geometry. Up to four sources can be used simultaneously. The target-to-substrate distance is 21\,cm and the inclination of the sources is 30$^\circ$. The substrate carrier can be heated to 1000$^\circ$C by an infrared heater from the backside. An electron beam evaporator with one crucible is placed in the center of the chamber at a distance of 50\,cm to the sample. The base pressure of the system was typically $5\,\cdot\,10^{-10}$\,mbar.

Elemental targets of Mn, Co, V, and Al of 99.95\,\% purity were used. The sputtering pressure was set to $2\cdot10^{-3}$\,mbar. The correct sputter power ratios were set up using a combined x-ray reflectivity and x-ray fluorescence technique.

All samples used in this study had the following stack sequence: MgO (001) single crystal / Mn$_{2-x}$Co$_x$VAl 18\,nm / Mg 0.5\,nm / MgO 1.5\,nm with $x\,=\,0\,/\,0.5\,/\,0.9\,/\,1.0\,/\,1.1\,/\,1.5\,/\,2$. The upper MgO was deposited by e-beam evaporation. The protective Mg / MgO bilayer was deposited after cooling the samples to prevent oxidation and interdiffusion. Diffraction measurements on Mn$_2$VAl films deposited at various temperatures revealed that a substrate carrier temperature of at least 600$^\circ$C was necessary to obtain good order, but temperatures above 700$^\circ$C lead to strong Mn sublimation, which can not be reliably compensated by higher sputtering power (compare with \cite{Kubota09}). Therefore all samples discussed in this paper were deposited at a carrier temperature of 700$^\circ$C.

X-ray diffraction (XRD), reflectometry (XRR), and fluorescence (XRF) were performed in a Philips X'Pert Pro MPD diffractometer with Cu anode, Bragg-Brentano and collimator point focus optics, an open Euler cradle and an Amptek fluorescence detector in a He enclosure.

X-ray magnetic circular dichroism (XMCD) was measured at beamline 6.3.1 of the Advanced Light Source. A magnetic field of $\pm$ 1.6\,T parallel to the incoming x-ray beam was applied, the sample surfaces were inclined by 30$^\circ$ with respect to the incoming beam. Element specific magnetic hysteresis loops were taken with a magnetic field of up to $\pm$ 2\,T. The magnetic field was switched for every energy point to obtain the dichroic signal. Data were taken at 20\,K, 150\,K, 200\,K, and 300\,K. All XMCD spectra were taken at least twice, with polarizations of +60\,\% and -60\,\%. Systematic measurements were performed in the surface sensitive total electron yield mode. Additionally, following an idea by Kallmayer \textit{et al.} \cite{Kallmayer07}, the visible light fluorescence of the MgO substrate was detected by a photo diode behind the sample. Thus, bulk information of the films could be obtained in x-ray transmission.

\subsection{Band structure calculations}
Band structure calculations of disordered compounds were performed with the \textit{Munich} SPRKKR package, a spin-polarized relativistic Korringa-Kohn-Rostoker code \cite{SPRKKR}. The ground state self-consistent potential calculations were performed on 834 \textbf{k} points in the irreducible wedge of the Brillouin zone. The exchange-correlation potential was approximated by the Perdew-Burke-Ernzerhof implementation of the generalized gradient approximation \cite{PBE}, the Fermi energy was determined using Lloyd's formula \cite{Lloyd72,Zeller08}. The angular momentum expansion was taken up to $l_\mathrm{max} = 3$. A scalar relativistic representation of the valence states was used in all cases, thus neglecting the spin-orbit coupling. For Mn$_2$VAl the atomic spheres approximation was applied and Co$_2$VAl was treated with full potential calculations. Half-metallic ground states were obtained for Mn$_2$VAl and Co$_2$VAl with their respective bulk lattice parameters. To account for disorder, the coherent potential approximation (CPA) was used. In our calculations with the ideally ordered L2$_1$ structure, Mn$_2$VAl has a total moment of 2.01\,$\mu_B$/f.u., with 1.54\,$\mu_B$ on Mn and -1.03\,$\mu_B$ on V. Co$_2$VAl has a total moment of 1.99\,$\mu_B$/f.u., with 0.87\,$\mu_B$ on Co and 0.28\,$\mu_B$ on V. These values are in good agreement with calculations presented by other authors \cite{Kandpal07}.

\section{Experimental results and discussion}

\subsection{Lattice structure}

\begin{figure}[t]
\begin{center}
\includegraphics[width=8.5cm]{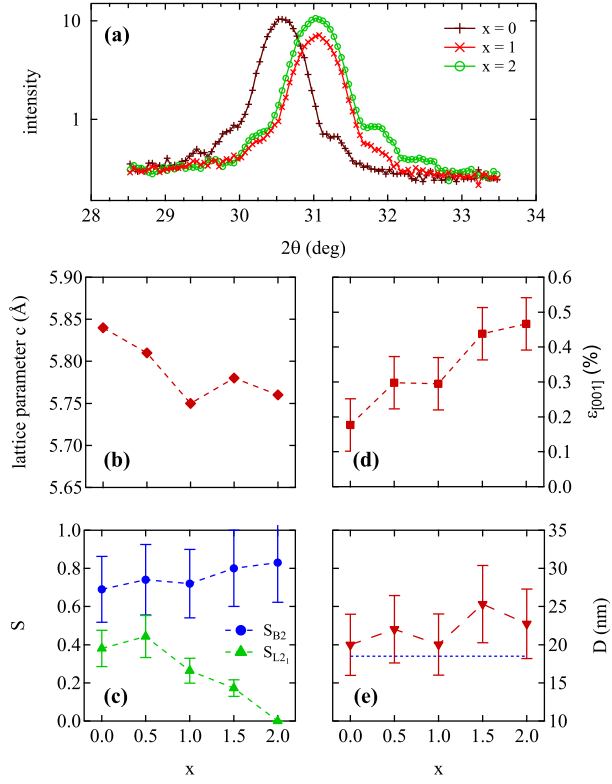}
\end{center}
\caption{\textbf{(a)}: $\theta$-$2\theta$ scans of the (002) reflections of Mn$_2$VAl ($x=0$) and Co$_2$VAl ($x=2$). Clear Laue oscillations are visible in both cases. \textbf{(b)}: out-of-plane lattice parameter $c$ as function of $x$. \textbf{(c)}: Order parameters $S_\mathrm{B2}$ and $S_{\mathrm{L2}_1}$ as functions of $x$. \textbf{(d)}: Microstrain $\varepsilon_{[001]}$ and \textbf{(e)}: coherence length $D$ and as functions of $x$. The dashed line in \textbf{(e)} denotes the film thickness.}
\label{xrd}
\end{figure}

All MCVA films were found to be highly epitaxial with MCVA [001] $\|$ MgO [001], rocking curve widths of 0.6$^\circ$ to 1.5$^\circ$, and an MCVA [100] $\|$ MgO [110] in-plane relation. Laue oscillations observed at the (002) reflections demonstrate the lattice and interface coherence of the films in the two limiting cases of Mn$_2$VAl and Co$_2$VAl (Fig. \ref{xrd}(a)). For $x=1$, however, the oscillations are less pronounced.

Figure \ref{xrd}(b) displays the out-of-plane lattice parameter $c$ as a function of $x$. According to Vegard's law \cite{Vegard21}, a linear decrease of the lattice parameter with increasing $x$ can be expected for a simple substitutional model. However, a significant deviation from this law is observed at $x=1$. This indicates, as we will see in detail later, a structural and magnetic order-disorder transition. For Mn$_2$VAl, $c$ is slightly lower than the bulk value of 5.875\,\AA{} \cite{Nakamichi83}; Co$_2$VAl has also a slightly reduced $c$ compared to the bulk value of 5.77\,\AA{} \cite{Ziebeck74}. This is compatible with a tetragonal distortion caused by the epitaxial matching with the substrate: the lattice is expanded in the plane and shrinks in the out-of-plane direction. For the case of Co$_2$TiSn we have recently performed first principles calculations of the change in total energy for this type of lattice distortion. In this case it is of the order of $25-50$\,meV, and is thus easily activated during the film growth \cite{Meinert10}. For the compounds presented here, we expect a similar energy range.

Takamura's extended order model for Heusler compounds \cite{Takamura09} was applied to obtain the order parameters $S_\mathrm{B2}$ and $S_{\mathrm{L2}_1}$ from the measured XRD peak intensities. Unlike Webster's model \cite{Webster71}, this model takes the dependence of $S_{\mathrm{L2}_1}$ on $S_\mathrm{B2}$ into account. The structure factors were obtained from the measured intensities by correcting for the Lorentz-Polarization term and the temperature factor with an effective Debye-Waller factor of $B_\mathrm{eff}=0.4$. $S_\mathrm{B2}$ is calculated from the four structure factor ratios of (002) and (222) versus (022) and (004), respectively. $S_{\mathrm{L2}_1}$ is calculated as the average of the (111) structure factor versus (022) and (004). The full atomic scattering factors including angular dependence and anomalous corrections were used in the numerical model calculations. As shown in Fig. \ref{xrd}(c), the Mn$_2$VAl films are ordered in the L2$_1$ structure with significant V-Al disorder ($S_{\mathrm{L2}_1} \approx 0.4$). With increasing Co content, the L2$_1$ order disappears in the alloy system; Co$_2$VAl does not show any sign of L2$_1$ ordering. On the other hand, the degree of B2 order increases slightly with increasing Co content, from $S_\mathrm{B2} = 0.7$ to $S_\mathrm{B2} = 0.8$, i.e., 85\,\% to 90\,\% of the Co atoms are on the 8c sites. However, we note here that disorder between Co, Mn, and V can not be identified with this method, because the atomic form factors are too similar.

A Williamson-Hall analysis \cite{Williamson53} of the integral peak widths of the (002), (004), and (006) reflections was performed. A Gaussian instrumental peak broadening and a Lorentzian convolution of grain size and strain effects were assumed, i.e., the contributions were separated with
\begin{equation}
B_\mathrm{obs}^2 = B_\mathrm{inst}^2 + B_\mathrm{ss}^2
\end{equation}
and
\begin{equation}
B_\mathrm{ss} \cdot \cos{\theta} = \frac{k\lambda}{D} + 4\, \varepsilon_{[001]}\, \sin{\theta}
\end{equation}
with $B_\mathrm{obs}$ being the observed width, $B_\mathrm{inst}$ the instrumental width, $B_\mathrm{ss}$ the size-strain width, the shape factor $k=0.9$, the coherence length (grain size) $D$ and the averaged $[001]$ component of the strain tensor $\varepsilon_{[001]}$. The analysis results are displayed in Fig. \ref{xrd}(d) and (e). The measured coherence length matches the film thicknesses quite well within the accuracy of the measuring and fitting procedure. A clear trend of increasing strain can be observed, from 0.18\,\% to 0.47\,\%. The lattice mismatch of Co$_2$VAl (3.1\,\%) is about 2.4 times as large as the mismatch of Mn$_2$VAl (1.3\,\%) with MgO. The same factor applies to the strain values, which verifies the high quality of the epitaxy. The lower degree of film coherence, the deviation from Vegard's law and the rather low strain in spite of the large lattice mismatch indicate an increased density of lattice defects in Mn$_{1}$Co$_{1}$VAl. The defects allow for relaxation of the film, which can reduce the microstrain at a loss of coherence.

Ziebeck and Webster found that Co$_2$VAl crystallizes in the L2$_1$ phase, but exhibits some preferential V-Al disorder \cite{Ziebeck74}. The samples measured by them were annealed at 800$^\circ$C for 24h. The samples by Kanomata \textit{et al.} were annealed at up to 1200$^\circ$C, and still exhibited a complex grain structure consisting of L2$_1$ and B2 ordered fractions. Deposition at 700$^\circ$C may thus be insufficient to promote L2$_1$ order in Co$_2$VAl. However, as stated initially, a higher deposition temperature was not usable because of Mn sublimation. 

\subsection{Magnetic and electronic structure}

\begin{figure*}[t]
\begin{center}
\includegraphics[scale=1]{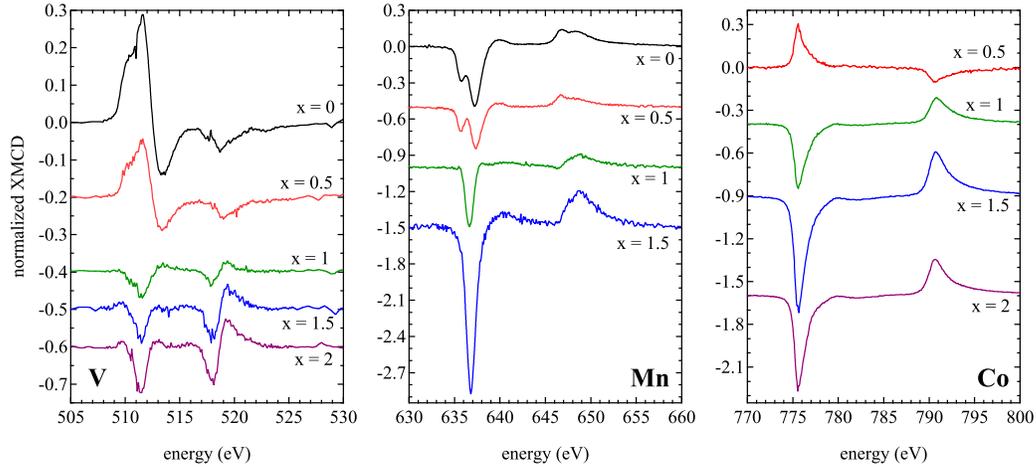}
\end{center}
\caption{Experimental XMCD spectra for V, Mn, and Co at 20\,K. The corresponding XAS spectra were normalized to a post-edge jump height of 1. The spectra for $x=0.9,1.1$ are similar to $x=1$ and are omitted for clarity. }
\label{xmcd_spectra}
\end{figure*}

\Table{\label{Tab:moments_corrected}Experimental total magnetic moments at 20\,K (given in $\mu_B$\,/\,f.u.) and estimated Curie temperatures.}
\br
		&				$m_\mathrm{tot}$	& $T_\mathrm{C}$	\\\hline\bs
Mn$_2$VAl					& 0.88			& $\gg$\,RT\\
Mn$_{1.5}$Co$_{0.5}$VAl		& 0.1				& - \\
Mn$_{1.0}$Co$_{1.0}$VAl		& 1.09			& $\approx 350$\,K\\
Mn$_{0.5}$Co$_{1.5}$VAl		& 2.29			& - \\
Co$_2$VAl					& 1.66			& $\approx 210$\,K\\
\br
\end{tabular}
\end{indented}
\end{table}

\begin{figure*}[t]
\begin{center}
\includegraphics[scale=1]{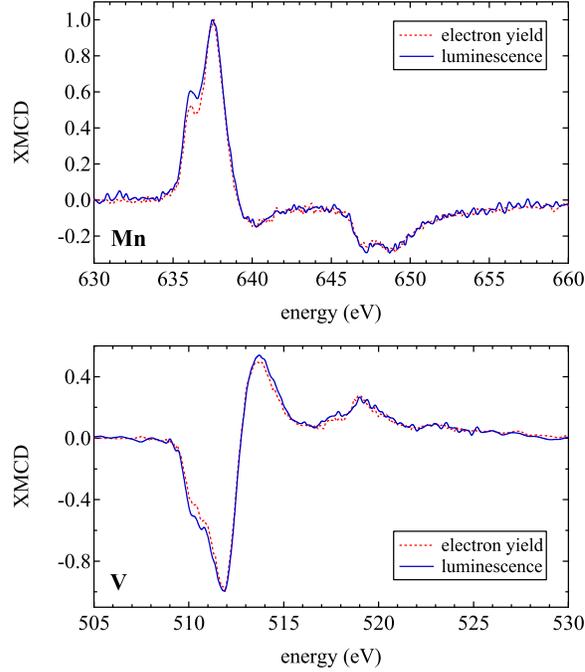}
\end{center}
\caption{Normalized XMCD spectra of Mn and V in electron yield and luminescence detection.}
\label{xmcd_TM_EY}
\end{figure*}

We begin with a discussion of the XMCD spectra in dependence on $x$, which are shown in Fig. \ref{xmcd_spectra} (a)-(c). For $x=0$, i.e., for pure Mn$_2$VAl, we find an antiparallel alignment of the Mn and V moments. This is preserved up to $x=0.5$, going along with an antiparallel coupling of Co to Mn. Here, we find the predicted ferrimagnetic order with the Co and V moments pointing opposite to the Mn moments. With further increasing $x$, all magnetic moments point in the same direction; the alloys become ferromagnets. This transition is closely related to chemical disorder which is indicated by the deviation of the lattice parameter from Vegard's law. Across the stoichiometry series the shape of the spectra changes significantly. Most prominently, the splitting of the V and Mn lines vanishes above $x=0.9$ and above. The appearance of this splitting is directly correlated with the appearance of ferrimagnetism. The line shape of the Mn XMCD for $x = 1.5$ is very similar to the Mn line shape in Co$_2$MnAl or Co$_2$MnSi \cite{Telling08}. For the ferrimagnetic coupling of Co and Mn, they have to be second nearest neighbors on octahedral positions. Co and Mn on tetrahedral nearest-neighbor positions couple ferromagnetically, as in Co$_2$MnGe \cite{Webster71} and the other Co$_2$Mn-based Heusler compounds.

To assert that the complex shape of the Mn and V spectra is not a surfacial effect, we have measured the transmitted x-ray intensity in luminescence detection at room temperature for Mn$_2$VAl. The XMCD spectra are almost equal in total electron yield and in transmission (see Figure \ref{xmcd_TM_EY}), although in both cases the L3 pre-peak is more pronounced in transmission. However, compared to the total area of the peaks, this deviation is small. The fine structure of the spectra is consequently related to the electronic structure of the films rather than to a surface effect.

\begin{figure*}[t]
\begin{center}
\includegraphics[scale=1]{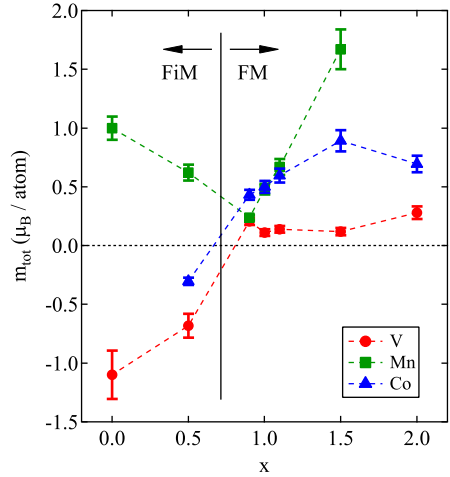}
\end{center}
\caption{Element specific magnetic moments as functions of $x$. Ferrimagnetic (FiM) order is observed for $x\leq 0.5$, ferromagnetic (FM) order is observed for $x\geq 0.9$.}
\label{xmcd_vs_x}
\end{figure*}

Using the sum rule analysis we extracted the spin and orbital magnetic moments from the XMCD spectra \cite{Chen95}. Table \ref{Tab:moments_corrected} summarizes the total magnetic moments obtained from sum rule analysis and provides estimates of the Curie temperatures obtained from temperature dependent XMCD for $x = 0, 1, 2$. Figure \ref{xmcd_vs_x} displays the element specific total moments in dependence on $x$. Because of dynamical screening effects of the x-ray field, the sum rules fail for the early 3d transition metals \cite{Ankudinov03}. To compensate the resulting spectral mixing effects, the apparent spin magnetic moments can be multiplied with correction factors as suggested by D\"urr \textit{et al.} and Scherz \textit{et al.}, i.e. $1.5$ for Mn \cite{Duerr97} and 5 for V \cite{Scherz02}. Actually, the applied correction factors depend on the actual electronic structure and can not be simply transferred to different systems. However, we assume that this influence is rather small, so that quantitative results can be obtained.

In Mn$_2$VAl we find a lowered Mn moment ($1\,\mu_B$) and an enhanced V moment ($-1.1\,\mu_B$), resulting in a total magnetization of $0.88$\,$\mu_B$/f.u. No change of the magnetic moments was observed at RT as compared to 20\,K, hence the Curie temperature is much higher than RT. The film is not well described by a pure L2$_1$ order model. As discussed earlier, the film has some disorder between Mn and (V,Al). In this case, Mn atoms reside on sites surrounded by other Mn atoms, which couple antiferromagnetically at short distance. Indeed, by calculating the self-consistent potential in SPRKKR with 20\% Mn-Al or Mn-V exchange, we find antiparallel coupling of the antisites. For Mn-Al exchange, the Mn(8a) moment is reduced to $1.22$\,$\mu_B$ and the Mn on the Al site has $-2.48$\,$\mu_B$. The V moment is reduced to $-0.83$\,$\mu_B$. This results in a total magnetization of $0.85$\,$\mu_B$/f.u., and the average Mn moment is consequently $0.85$\,$\mu_B$. In the case of Mn-V exchange, the Mn(8a) moment remains at $1.58$\,$\mu_B$ and the Mn on the V site has $-2.63$\,$\mu_B$. The V moment on the 4b site is $-0.87$\,$\mu_B$ and $+0.84$\,$\mu_B$ on the 8a site. In this case the total moment is $1.78$\,$\mu_B$/f.u., with an average Mn moment of $1.16$\,$\mu_B$. Further, the case of Mn-Al exchange is energetically preferred with respect to the Mn-V exchange. Seeing the low total and Mn moments and the high V moment, a preferential Mn-Al exchange in Mn$_2$VAl is thus in good agreement with the structural and the magnetic data. Our calculations show that the 20\,\% Mn-Al disorder and B2 disorder barely influence the half-metallic gap of Mn$_2$VAl. For B2 disorder, the total magnetic moment also remains unaffected. In contrast, 20\,\% Mn-V disorder destroy the gap. This is in contrast to the findings by Luo \textit{et al.}, obtained with a supercell approach in a pseudopotential code. They state that the gap is preserved under 25\,\% Mn-V disorder \cite{Luo08}.

Co$_2$VAl has a reduced Co moment ($0.69\,\mu_B$) and a V moment of $0.28\,\mu_B$, giving a total magnetization of 1.66\,$\mu_B$/f.u. The film has B2 order, which is expected to reduce the magnetization from the highly ordered L2$_1$ case.  We find magnetic moments of 0.75\,$\mu_B$ for Co and 0.4\,$\mu_B$ for V in a B2 ordered SPRKKR calculation, with a total moment of 1.86\,$\mu_B$/f.u., in good agreement with our measurements. Some additional disorder involving Co and V could explain the further reduced moments. The Curie temperature is about 210\,K, which is significantly lower than the value for bulk samples (310\,K \cite{Ziebeck74}). A calculation of the Curie temperature with SPRKKR within the mean field approximation (see \cite{Meinert11} for details of the procedure) yields 352\,K in the L2$_1$ case and 165\,K in the B2 ordered case. The observed significant reduction of the Curie temperature in the disordered alloy is thus in agreement with theory. The half-metallic gap of Co$_2$VAl vanishes in the B2 structure.

At $x=0.5$, a nearly complete magnetic compensation with a total moment of only $0.1$\,$\mu_B$/f.u. is observed. Remarkably, at $x=1.5$ the total magnetic moment becomes larger than 2\,$\mu_B$/f.u., caused by the high Mn moment of 1.67\,$\mu_B$. This is in agreement with the different Mn line shape: in, e.g., Co$_2$MnAl, in which Mn has a similar line shape, Mn has a moment of about 3\,$\mu_B$ \cite{Webster71}. Thus, the mechanism mainly responsible for the ferromagnetic coupling of all moments is the preferentially tetrahedral (instead of octahedral) coordination of Mn atoms with Co.
\section{Conclusions}

Epitaxial thin films of Mn$_{2-x}$Co$_x$VAl have been synthesized on MgO (001) substrates by DC and RF magnetron co-sputtering. It was intended to observe a ferrimagnetic compensation of the magnetization at $x=1$. The films have significant chemical disorder, depending on the degree of Mn-Co substitution. Mn$_2$VAl was found to be L2$_1$ ordered, with a preferential Mn-Al disorder and additional V-Al disorder. The Mn-Al disorder reduces the total moment considerably, because the nearest-neighbor Mn atoms couple antiferromagnetically in this configuration. Accordingly, the magnetization of Mn$_2$VAl is very sensitive to disorder involving Mn. However, the band structure calculations suggest that only Mn-V disorder has an influence on the half-metallic gap.

Because of the disorder, a nearly complete magnetic compensation was observed for Mn$_{1.5}$Co$_{0.5}$VAl. With further Co substitution, the electronic structure changes considerably, and a parallel coupling of Co, Mn, and V was observed. We suppose that Co and Mn become preferentially nearest neighbors, which leads to a parallel coupling of their magnetic moments.

The Co$_2$VAl films, being the second extremum of the substitutional series, had B2 order. The band structure calculations with B2 order suggest reduced moments, but the experimentally determined moments are further reduced, which indicates additional disorder involving Co. The Curie temperature was significantly reduced, which is in agreement with the trend observed in the mean field calculation. It is in principle possible to obtain a high degree of L2$_1$ order in bulk Co$_2$VAl by appropriate thermal treatment, but our maximum substrate temperature was limited by Mn evaporation. While it may be possible to obtain the correct occupation for the ferrimagnetic compensation in the bulk, it seems not possible to obtain films with a high degree of order.

\section*{Acknowledgements}
The authors gratefully acknowledge financial support by the Deutsche Forschungsgemeinschaft (DFG)  and the Bundesministerium f\"ur Bildung und Forschung (BMBF). They thank for the opportunity to work at BL 6.3.1 of the Advanced Light Source, Berkeley, USA, which is supported by the Director, Office of Science, Office of Basic Energy Sciences, of the U.S. Department of Energy under Contract No. DE-AC02-05CH11231.

\end{document}